\documentclass[fleqn,12pt]{article} %{epsconf}
\usepackage{graphicx}
\usepackage{epsfig} % use this package to include EPS format figures
\usepackage{wrapfig}
\usepackage{amsmath}
\begin{document}

\begin{center}
\Large \bf {Derivation and quantitative analysis of the differential
self-interrogation Feynman-alpha method}
\end{center}
\vspace{0.5cm}
%\newline
%\newline
\begin{center}
{Johan Anderson\footnote{johan@nephy.chalmers.se}, L\'en\'ard P\'al$^{2}$, Imre P\'{a}zsit$^{1,3}$, Dina Chernikova$^{1}$ and Sara Pozzi$^3$ }
\end{center}
\begin{center}
$^1$ Department of Nuclear Engineering, Chalmers University of Technology, SE-41296, G\"{o}teborg, Sweden \\
\end{center}
\begin{center}
$^{2}$ KFKI Atomic Energy Research Institute H-1525 Budapest 114, POB 49, Hungary
\end{center}
\begin{center}
$^{3}$ Department of Nuclear Engineering and Radiological Sciences, University of
Michigan, Ann Arbor Michigan, USA
\end{center}

%\pacs{05.40.-a}
%\pacs{05.70.Ln}
%\pacs{68.35.Fx}

\begin{abstract}
\noindent A stochastic theory for a branching process in a neutron
population with two energy levels is used to assess the
applicability of the differential self-interrogation Feynman-alpha
method by numerically estimated reaction intensities from
Monte Carlo simulations. More specifically, the variance to mean or
Feynman-alpha formula is applied to investigate the appearing
exponentials using the numerically obtained reaction intensities.
\end{abstract}

\section{Introduction}
Energy dependent aspects of neutron counting have been used to
detect and determine the content of nuclear materials for several
decades. The prime application is the Differential Die-Away Analysis
(DDAA) method \cite{kunz1982, croft2003, jordanphd, jordan2007,
jordan2008}. It is a deterministic method in which the time
dependence of the detection rate of fast neutrons, after that an
initial pulse of fast neutrons is injected to the sample, is used to
determine the presence of fissile materials such as $^{235}U$ and
$^{239}Pu$. Its application is suitable when the fissile material is
embedded in a moderating surroundings such that the source neutrons
induce thermal fission after having slowed down.

The drawback of pulsed measurements such as the DDAA method is the necessity of
the use of a neutron generator. The stochastic generalization of the
DDAA method was recently suggested as an alternative method which eliminates 
the need for such an external source, namely the so-called Differential Die-away Self-Interrogation
(DDSI) technique \cite{menlove09}. The DDSI method utilizes 
the inherent spontaneous neutron
emission of the sample. In the absence of a trigger
signal, the temporal decay of the correlations as
a function of the time delay between two detections of fast neutrons
is used in the DDSI method. This corresponds to a Rossi-alpha measurement with two
energy groups.

In Ref. \cite{menlove09} the functional form of the DDSI formula,
i.e. the two-group Rossi-alpha formula, was derived in an empirical
way. This formula was therefore later derived from first principles
with the use of backward-type Kolmogorov equations \cite{a1,a2}. A
two-group theory of the Rossi- and Feynman alpha formulae is interesting
also in areas other than nuclear safeguards, such as pulsed and
stationary source driven experiments measuring the reactivity in
fast cores of accelerator driven sub-critical systems. In such
experiments it was found that two exponentials appear, indicating
that the temporal behavior of the fast and thermal neutrons is
separated, especially in fast reflected cores. This amplifies the
need for the two group versions of the Feynman and Rossi-alpha
formulae \cite{a1,a2}.

In this contribution, a stochastic theory for a branching process in
a neutron population with two energy levels is investigated based on
the previous results of Refs \cite{a1,a2,anderson2011}. In
particular, a counterpart of the DDSI formula, the Differential
Self-interrogation Variance to Mean or Feynman-alpha formula, which
we shall call the DSVM formula, is derived by using the master equation or Kolmogorov
forward approach. The model includes a spontaneous fission source of
fast neutrons, absorption in both groups, down-scattering (removal)
from the fast to the thermal group, thermal fission, and detection
of fast neutrons.

Practical applicability of the method, which is based on the
observation of two different exponentials, depends on the
quantitative values of the various within- and inter-group neutron
reaction intensities, which appear as coefficients in the equations.
This is in contrast to the energy-independent theory of multiplicity
where the only appearing parameter is the first collision
probability. To assess the applicability and expected performance of
the DSVM method in practical situations, the above mentioned
reaction intensities were determined from numerical Monte Carlo
(MCNP4c) simulations. Significance of different values of the
reaction intensities of thermal and fast neutrons in the performance
of the method is discussed.

\section{A variance to mean formula for the detected fast neutrons}
In this section we will briefly discuss the fundamentals of the
derivation of the variance to mean or Feynman-alpha for the two
particle type system by using the Kolmogorov forward approach
following Ref. \cite{anderson2011}. In the model we have included a
compound Poisson source of fast neutrons described by the source
strength $S_1$ that releases $n$ particles with probability $p_q(n)$
at an emission event (i.e. spontaneous fission). We have assumed
that the source is switched on at time $t=t_0 < 0$, although the
dependence on $t_0$ will not be denoted. The detection rate of
particles is included and is denoted by the intensity $\lambda_d$.
We will start by giving a presentation of the analytical model
consisting of a differential equation for the probability $P(N_1,
N_2, Z_1,t)$ for having $N_1$ fast, $N_2$ thermal neutrons at time $t$ in the system
and having detected $Z_1$ fast particles in the interval $(0,t)$. In deriving
this differential equation we have summed all mutually exclusive
events during an infinitesimally small time interval $dt$ and we
find for the probability $P(N_1,N_2, Z_1,t)$ the differential equation
\begin{eqnarray}\label{eq:1.1}
\frac{\partial P(N_1,N_2,Z_1,t)}{\partial t} & = & - (\lambda_1 N_1 + \lambda_2 N_2 + S_1)P(N_1, N_2, Z_1,t) \nonumber \\
& + & \lambda_{1a} (N_1 + 1) P(N_1 + 1, N_2, Z_1,t) \nonumber \\
& + & \lambda_{2a} (N_2 + 1) P(N_1, N_2 + 1, Z_1,t) \nonumber \\
& + & \lambda_R (N_1 + 1) P(N_1 + 1, N_2 - 1, Z_1, t) \nonumber \\
& + & \lambda_{2f} (N_2 + 1) \sum_k^{N_1} f(k) P(N_1 - k, N_2 + 1, Z_1,t) \nonumber \\
& + & \lambda_d (N_1 + 1) P(N_1 + 1, N_2, Z_1 - 1,t) \nonumber \\
& + & S_1 \sum_n^{N_1} p_q(n) P(N_1 - n, N_2, Z_1,t).
\end{eqnarray}
Here, $\lambda_1$ and $\lambda_2$ are the decay constants (total
reaction intensities) for fast and thermal particles, whereas
$\lambda_{1a}$, $\lambda_{2a}$ are the absorption (actually,
capture) intensities of fast and thermal particles, respectively.
The removal of fast particles into the thermal group is described by
$\lambda_R$ while fission resulting from the thermal particles
happens with the intensity of $\lambda_{2f}$. The total intensities
are given by
\begin{eqnarray}\label{eq:1.2}
\lambda_1 = \lambda_{1a} + \lambda_R + \lambda_d,
\end{eqnarray}
and
\begin{eqnarray}\label{eq:1.3}
\lambda_2 = \lambda_{2a} + \lambda_{2f}.
\end{eqnarray}
We derive the equations for the factorial moments by using the
generating function of the form
\begin{eqnarray}\label{eq:1.4}
G(X,Y,Z,t) & = & \sum_{N_1} \sum_{N_2} \sum_{Z_1} X^{N_1} Y^{N_2} Z^{Z_1} P(N_1,N_2,Z_1,t),
\end{eqnarray}
and describe the time evolution of the process by a partial
differential equation in terms of the generating function as,
\begin{eqnarray}\label{eq:1.5}
\frac{\partial G}{\partial t} & = & (\lambda_{1a}  + \lambda_R Y + \lambda_d Z -
\lambda_1 X) \frac{\partial G}{\partial X} + (\lambda_{2a}  + \lambda_{2f} q(X) -
\lambda_2 Y) \frac{\partial G}{\partial Y} \nonumber \\
& + & S_1 (r(X) - 1) G,
\end{eqnarray}
where
\begin{eqnarray}
q(X) & = & \sum_k f_k X^k, \label{eq:1.61} \;\;\;\;\;\;\;\;\;\;  \mbox{and} \\
r(X) & = & \sum_n p_q(n) X^n. \label{eq:1.62}
\end{eqnarray}
Here, $f_k$ is the probability of having exactly $k$ neutrons
produced in an induced fission event. We have used the definition of
the derivatives of the expressions (\ref{eq:1.61}) and (\ref{eq:1.62})
as $\nu_1 = d q/dX|_{X=1}$ and $r_1 = d h/dX|_{X=1}$, which stand for the expectations of the number of neutrons from induced and sponaneous fissions, respectively \cite{pazpal08}. We note that
for $t_0 \rightarrow -\infty$, the expectations of fast neutrons
($\langle N_1 \rangle$) and  thermal neutrons ($\langle N_2
\rangle$) will reach steady state due to the stationary source term
with intensity $S_1$. The solutions to the system of differential
equations are found by differentiation of equation (\ref{eq:1.5}) with
respect to ($X,Y,Z$) and then letting ($X = Y = Z = 1$). These
read as
\begin{eqnarray}
\langle N_1 \rangle & = & \bar{N}_1 = \frac{\lambda_2 S_1 r_1}{\lambda_1
\lambda_2 - \nu_1 \lambda_R \lambda_{2f}} =  \frac{\lambda_2 S_1 r_1}{\omega_1 \omega_2}, \label{eq:1.10}\\
\langle N_2 \rangle & = & \bar{N}_2 = \frac{\lambda_R S_1 r_1}{\omega_1 \omega_2}, \label{eq:1.11} \\
\langle Z_1 \rangle & = & \varepsilon \lambda_{2f} \bar{N}_1 t, \label{eq:1.12}
\end{eqnarray}
where $\varepsilon = \lambda_d/\lambda_{2f}$ and we have used the additional definitions $\omega_1$ and $\omega_2$ and $ \nu_{eff} $  as
\begin{eqnarray}
-\omega_1 & = & -\frac{1}{2}(\lambda_1 + \lambda_2) + \frac{1}{2}
\sqrt{(\lambda_1 - \lambda_2)^2 + 4 \lambda_1  \lambda_2 \nu_{eff}} ,\label{eq:1.13} \\
-\omega_2 & = & -\frac{1}{2}(\lambda_1 + \lambda_2) - \frac{1}{2}
\sqrt{(\lambda_1 - \lambda_2)^2 + 4 \lambda_1  \lambda_2 \nu_{eff}},\label{eq:1.14} \\
\nu_{eff} & = & \nu_1 \frac{\lambda_R \lambda_{2f}}{\lambda_1 \lambda_2}. \label{eq:1.15}
\end{eqnarray}
It can  be noted that in reactor physics terminology, $\nu_{eff}$ is identical with $k_{eff}$ of a multiplying system and that the expectation of the detections increases linearly
with time. This is due to the fact that it is determined by
integration of the expectation of the neutron number with respect to time.
In order to find the variance of
the detector counts we need to determine the second moments by yet
another differentiation with respect to ($X,Y,Z$) followed by
letting ($X = Y = Z = 1$). We find the variance of the detector
counts through the relation $\sigma_Z^2 = \langle Z_1 \rangle +
\mu_{ZZ}$ where the modified variance $\mu_{ZZ}$ is defined as
$\mu_{ZZ} = \langle Z(Z-1) \rangle - \langle Z\rangle^2 =
\sigma_{ZZ}^2 - \langle Z\rangle$ while in general we have $\mu_{X
Y} = \langle X Y\rangle - \langle X \rangle \langle Y \rangle$. The
differentiation procedure results in a system of six ordinary
differential equations for the second order modified moments. 
The differentiation procedure gives a system of six dynamical equations of the modified second moments as
\begin{eqnarray}
\frac{\partial}{\partial t} \mu_{X X} & = & - 2 \lambda_1 \mu_{X X} + 2 \nu_1 \lambda_{2 f} \mu_{X Y}  + \nu_2 \lambda_{2f} \bar{N}_2 + S_1 r_2, \label{eq:1.16} \\
\frac{\partial}{\partial t} \mu_{X Y} & = & -(\lambda_1 + \lambda_2) \mu_{X Y} + \lambda_R \mu_{X X} + \nu_1 \lambda_{2 f} \mu_{Y Y}, \label{eq:1.17}\\
\frac{\partial}{\partial t} \mu_{Y Y} & = & - 2 \lambda_2 \mu_{Y Y} + 2 \lambda_R
 \mu_{X Y},\label{eq:1.18} \\
\frac{\partial}{\partial t} \mu_{Z X} & = & - \lambda_1 \mu_{Z X} + \nu_1 \lambda_{2f} \mu_{Z Y} + \lambda_d \mu_{X X},\label{eq:1.19} \\
\frac{\partial}{\partial t} \mu_{Z Y} & = & - \lambda_2 \mu_{Z Y} + \lambda_R \mu_{Z X}  + \lambda_d  \mu_{X Y}, \label{eq:1.20} \\
\frac{\partial}{\partial t} \mu_{ZZ} & = & 2\epsilon \lambda_{2f} \mu_{X Z}, \label{eq:1.21}
\end{eqnarray} 
The equation system and its solution is rather analogous to the case of the Feynman-alpha equations with one neutron energy group but including delayed neutrons, as given in Ref. \cite{pazsit1999}. Although an analytical solution for the general time-dependent system of equations (\ref{eq:1.16}) - (\ref{eq:1.21}) would be hard to find, we note that in the stationary state the system breaks down into two systems such that the solution of the first three equations is independent from the second, such that the moments  $\langle \mu_{X X} \rangle = \bar{\mu}_{XX} $, $\langle \mu_{X Y} \rangle = \bar{\mu}_{XY} $ and $\langle \mu_{Y Y} \rangle = \bar{\mu}_{YY} $ are constants. The equations describing detected particles need to be solved retaining the full time evolution by e.g. Laplace transforms. Moreover, it is found that the sought moment $\langle \mu_{Z Z} \rangle $ is determined by quadrature of moment $\langle \mu_{X Z} \rangle$. We find the constant 2nd moments as,
\begin{eqnarray}
\bar{\mu}_{XX} & = & \frac{(\lambda_2^2 + \omega_1 \omega_2)
(\nu_2 \lambda_{2f} \bar{N}_2 + S_1 r_2)}{2(\lambda_1 + \lambda_2)\omega_1 \omega_2}, \label{eq:1.22}\\
\bar{\mu}_{XY} & = &  \frac{\lambda_2 \lambda_R (\nu_2 \lambda_{2f}
\bar{N}_2 + S_1 r_2)}{2(\lambda_1 + \lambda_2)\omega_1 \omega_2}, \label{eq:1.23} \\
\bar{\mu}_{YY} & = & \frac{\lambda_R^2 (\nu_2 \lambda_{2f}
\bar{N}_2 + S_1 r_2)}{2(\lambda_1 + \lambda_2)\omega_1 \omega_2}. \label{eq:1.24}
\end{eqnarray}
Here we have used the notations $\nu_2 = d^2 q/dX^2|_{X=1}$ and $r_2 = d^2 h/dX^2|_{X=1}$ for the second factorial moments \cite{pazpal08}. The objective now is to solve (\ref{eq:1.19}) and (\ref{eq:1.20}) by Laplace transform methods and we find the transformed identity as,
\begin{eqnarray}
\tilde{\mu}_{XZ} = \frac{\nu_1 \lambda_d \lambda_{2f} \bar{\mu}_{XY}}{s H(s)} + \frac{(s+\lambda_2)\lambda_d \bar{\mu}_{XX}}{s H(s)} \label{eq:1.25}
\end{eqnarray}
with
\begin{eqnarray}
H(s) = s^2 + (\omega_2 + \omega_1)s + \omega_1 \omega_2. \label{eq:1.26}
\end{eqnarray} 
Note that we have assumed that the initial values of the moments $\mu_{XZ}$ and $\mu_{YZ}$ were equal to zero at $t=0$ (at the start of the measurement), hence the roots of $H(s)$ determine the temporal behavior of the Feynman-alpha formula. Moreover, the solution has many similarities to that found in Ref. \cite{pazsit1999}. We determine the variance to mean or Feynman-alpha formula by utilizing the relation for the variance $\sigma_{ZZ} = \langle Z \rangle + \mu_{ZZ}$ and after some algebra we find,
\begin{eqnarray}
\frac{\sigma_{ZZ}(T)}{\langle Z_1 \rangle} = 1 + Y_1 (1 - \frac{1-e^{- \omega_1 T}}{\omega_1 T}) + Y_2 (1 - \frac{1-e^{- \omega_2 T}}{\omega_2 T}). \label{eq:1.27}
\end{eqnarray}
Here, the complete expressions for $Y_1$ and $Y_2$ are quite
lengthy. However, it turns out that the sum $Y_0 = Y_1 + Y_2$ takes a
rather simple form that also determines the value of the
Feynman-alpha for large measurement times $T \rightarrow \infty$ as,
\begin{eqnarray}
Y_0 =  Y_1 + Y_2 = \nu_2 \frac{\lambda_d \lambda_2 \lambda_R \lambda_{2f}}
{\omega_1^2 \omega_2^2}.\label{eq:1.28}
\end{eqnarray}
In the next section we will consider some quantitative examples of the
variance to mean formula in Equation (\ref{eq:1.27}).

\section{Quantitative assessment of the variance to mean formula}
We will now discuss the quantitative application of the variance to mean formula (\ref{eq:1.27}) for a MOX fuel assembly, considering setups when the MOX fuel was homogeneously mixed with a different amount of polyethylene moderator. The fuel sample was in form of a sphere which has the same amount of material as one single fuel pin, yielding a radius of 3.36 cm, of typical MOX fuel (U(94.14\%)O2 + Pu(5.86\%)O2) with approximately 4\%  fissile content. The amount of moderator mixed into the fuel was taken as that corresponding to an outer reflector shell of 4 different thicknesses: 5, 10, 15 and 20 cm. In all cases the neutron source was assumed in the sample centre. In addition, a case was considered where the source was assumed to be equally distributed in the volume (designated as VDS, Volume Distributed Source) with the amount of moderator corresponding to 5 cm thickness. The geometry of the calculations is displayed in the r.h.s. of Figure 1, whereas the l.h.s. of the Figure shows the equivalent heterogeneous model, displaying a surrounding moderator which was mixed homogeneously in the sample for the simulations. We have hence carried out five Monte-Carlo (MCNP4с) simulations \cite{breis} in order to estimate the reaction intensities in the spent fuel assemblies with varying amount of moderation where the radius is taken as fissile material + moderator ($3.36 + d$ cm). In the simulations we have assumed a two energy group approximation with a $0.46$ eV energy cut-off. Moreover, the energies of the source neutrons were sampled from the Watt fission spectrum, originating from the spontaneous fission of $^{240}Pu$. The energy spectrum is determined by the function,
\begin{eqnarray}
p(E) \sim e^{-E/a} \sinh (b E)^{1/2}
\end{eqnarray}
where $a = 0.799$ MeV and $b = 4.903$ $MeV^{-1}$.

\begin{table}[ht]
\caption{Simulation Results} % title of Table
\centering % used for centering table
\begin{tabular}{c c c c c c c} % centered columns (5 columns)
\hline\hline %inserts double horizontal lines
Intensity & 5 cm & 10 cm & 15 cm & 20 cm & VDS \\ [0.5ex] % inserts table
%heading

$\lambda_1$ $[1/s]$     & 1.560 & 1.230 & 0.625 & 0.185 & 1.490      \\
$\lambda_2$ $[1/s]$     & 0.637 & 1.220 & 1.260 & 1.180 & 0.325      \\
$\lambda_R$ $[1/s]$     & 0.450 & 0.621 & 0.391 & 0.094 & 0.452      \\
$\lambda_{2f}$ $[1/s]$  & 0.243 & 0.285 & 0.171 & 0.094 & 0.118      \\

\end{tabular}
\label{table:simulat} % is used to refer this table in the text
\end{table}

\begin{figure}[ht!]
\includegraphics[width=12cm, height = 5.5cm]{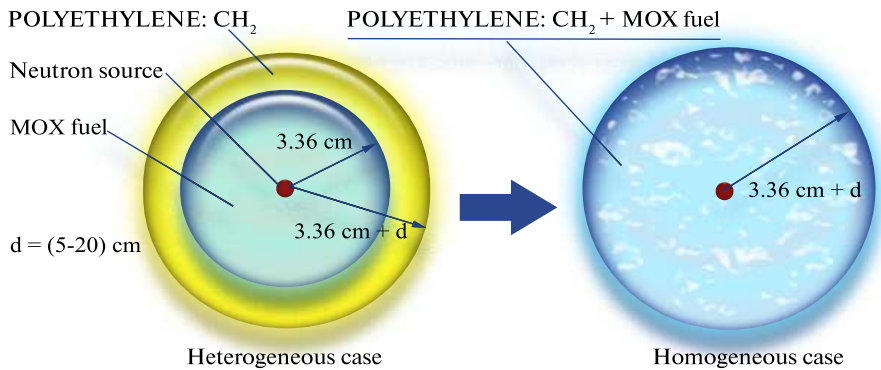} \label{fig:fig1}
\caption{Show the geometry used for the Monte Carlo (MCNP4c) simulations for the homogenized cases (left figure showing fuel mixed with different amounts of polyethylene moderator) compared to the heterogeneous cases (right).}
\end{figure}

\begin{figure}[ht!]
\begin{minipage}[b]{0.5\linewidth}
\centering
\includegraphics[width=7cm, height = 7cm]{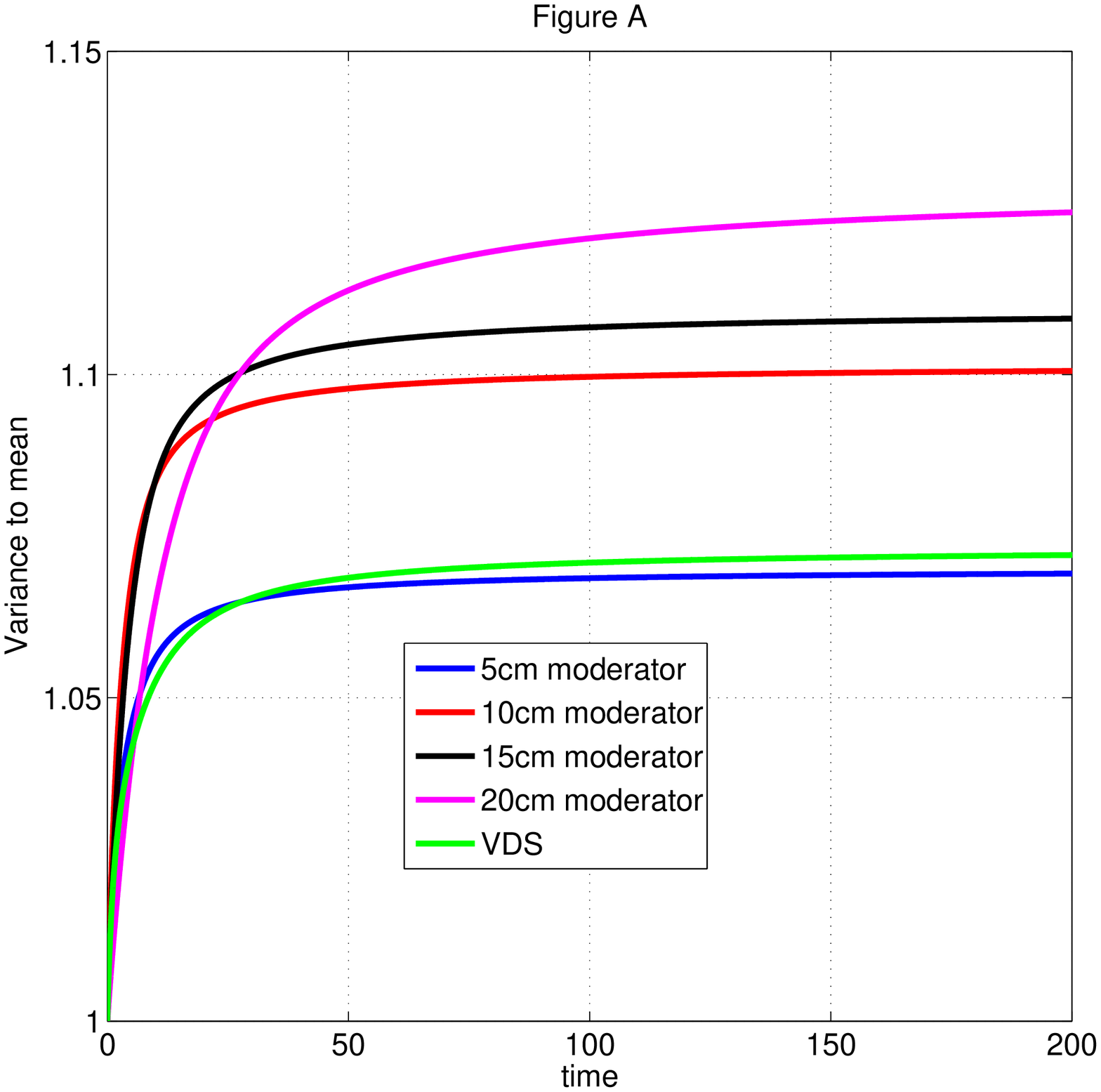}
%\caption{default}
\end{minipage}
\hspace{0.5cm}
\begin{minipage}[b]{0.5\linewidth}
\centering
\includegraphics[width=7cm, height = 7cm]{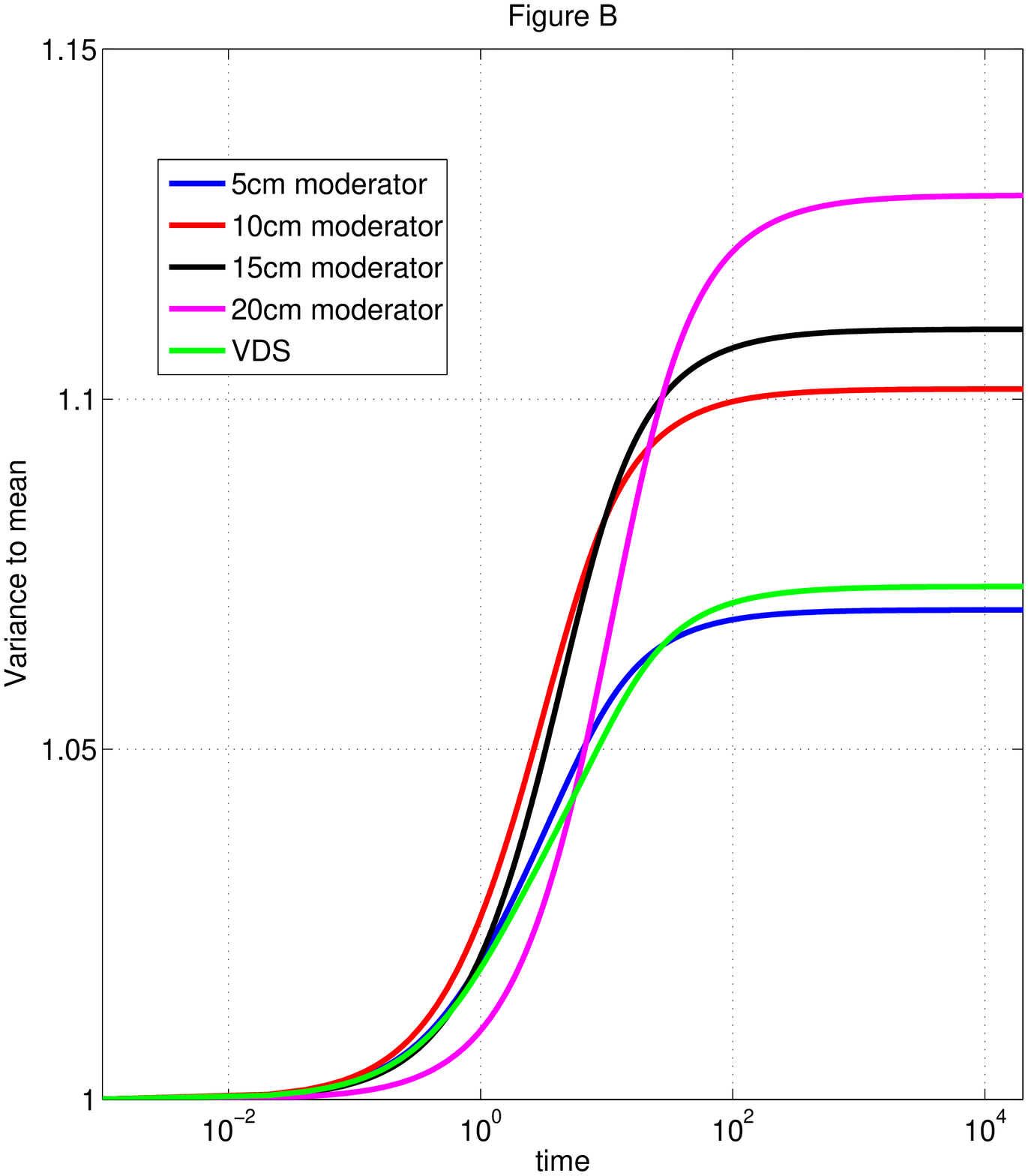}
%\caption{default}
\end{minipage}
\caption{(A and B) The Feynman-alpha formula is shown for the cases
generated by Monte Carlo (MCNP4c) simulations with 5-20 cm moderator
in lin-lin scale (Figure A) and lin-log scale (Figure B).}
\label{fig:fig2}
\end{figure}

The simulation results are shown
in Table \ref{table:simulat} for the cases described above.
The reaction intensities obtained from the simulations were normalized
to one starting neutron and are obtained with a relative error less than 1\%. The resulting
reaction intensities are used in the variance-to-mean expression in
Eq. (\ref{eq:1.27}) and the results are displayed in Figures 2 A and B. In Figure 2A, the variance to mean in lin-lin scale is shown for the parameters in Table 1 whereas the remaining parameters are $\nu_1 = 2.87$, $\nu_2 = 4.635$, $\lambda_d = 0.1$, $r_1 = 1.0$ and $r_2 = 0.0$. We find that that the difference between the variance to mean for the volume distributed source (VDS in table
(\ref{table:simulat}), green line) and the point source with 5 cm
moderator (blue line) is very small indicating that the point source model 
is sufficient for describing the system. We note that the values of $\nu_{eff}$ are 0.316, 0.338, 0.244, 0.116 and 0.316 for the cases with 5 cm, 10 cm, 15 cm, 20 cm moderator and the VDS, respectively. In Figure 2B the variance to mean from Figure 2A is displayed in
lin-log scale using the same parameter values as in Figure 2A. The reaction intensities of the moderated cases are approximately of the same order of magnitude and thus the two exponentials are not easily distinguished.

\section{Discussion and conclusions}
We have developed a forward Kolmogorov approach for the two group
theory of the Differential Die-Away Self Interrogation Variance to Mean (DSVM), including a compound Poisson source and the detection process. The results agree with those
calculated by the backward approach as reported in \cite{a1, a2}. We
have used Monte Carlo simulations to find the reaction intensities
needed to quantitatively assess the DSVM formula. We find that, unlike in the DDSI method (i.e.
the two-group version of the Rossi-alpha method), the presence of
two exponents in the solution is most often not clearly visible. This means that detection of the presence of fissile material may not be as obvious as with the Rossi-alpha method. By
using the two group Rossi-alpha expression 4.34 in Ref. \cite{a1},
we find the reverse situation, where in most cases with moderator
the two exponentials appear. On the other hand, the
determination of the exponents $\omega_1$ and $\omega_2$ by curve
fitting could be more accurate in certain cases than with the DDSI
method. Elucidating on the diagnostic value of the exponents in
terms of determination of the sample parameters is not clear yet,
and it requires further investigations, which will be reported in
future work.
\section{Acknowledgements}
This work was supported by the Swedish Radiation Safety Authority (SSM).

\end{document}